# Anomaly Detection in Business Process Runtime Behavior – Challenges and Limitations

Kristof Böhmer
University of Vienna
Faculty of Computer Science
Email: kristof.boehmer@univie.ac.at

Stefanie Rinderle-Ma
University of Vienna
Faculty of Computer Science
Email: stefanie.rinderle-ma@univie.ac.at

*Abstract*—Anomaly detection is generally acknowledged as an important problem that has already drawn attention to various domains and research areas, such as, network security. For such "classic" application domains a wide range of surveys and literature reviews exist already – which is not the case for the process domain. Hence, this systematic literature review strives to provide an organized holistic view on research related to business process runtime behavior anomaly detection. For this the unique challenges of the process domain are outlined along with the nature of the analyzed data and data sources. Moreover, existing work is identified and categorized based on the underlying fundamental technology applied by each work. Furthermore, this work describes advantages and disadvantages of each identified approach. Based on these information limitations and gaps in existing research are identified and recommendations are proposed to tackle them. This work aims to foster the understanding and development of the process anomaly detection domain.

## I. Introduction

Business processes integrate a wide range of data sources, applications, IT systems, and manage public but also confidential private information, cf. [1]. In addition, processes tend to be executed in flexible diverse infrastructures and interconnected environments where they fulfill viable roles and enable critical business operations. Hence, processes cannot only be beneficial for an organization but also for attackers that can exploit them, for example, to spread attacks over multiple IT systems and partners, cf. [2], [3]. Moreover, processes are not only applied in human driven scenarios but also in automatic execution environments – which are especially security critical. This is because automatic process executions are less strictly controlled and monitored by humans. Hence, attackers can explore potential attack vectors fast and automatically, likely, without ever being noticed, cf. [4]. These aspects show that processes are not only important but also a security critical aspect in today's organizations. Hence, applying security measures is a necessity to prevent security incidents which range from targeted or random outside attacks over malicious insiders to fraud that can harm an organization.

Furthermore organizations are also required by law to take measures which foster secure process executions. For example upcoming EU regulations, such as, the General Data Protection Regulation (GDPR) enforce significant fines for security incidents, i.e., up to €20 million or 4% of the attacked organization's global annual turnover for the preceding financial year, cf. [5]. The GDPR is not the only regulation that encourages process based organizations to implement security measures. Another example is the Health Insurance Portability and Accountability Act (HIPAA) which was enacted in 1996. It establishes a U.S. wide standard for electronic *health care data* transmissions, storage, and processing. It states, for example, in §1173(d)(2)(AB) that each entity needs to ensure (A) the integrity and confidentiality of private data; (B) to protect private data against reasonable anticipated threats or hazards to the security or integrity of the information; and (C) unauthorized uses or disclosure of information. This law applies to each entity that handles health care information, i.e., a wide range of entities is affected from hospitals, over specialists, to medical laboratories. Similar regulations – with even broader implications – can be found in many areas, such as, the 44 U.S. Code §3542 (2012), to give only one additional example. Hence, we assume that the monitoring and control of process executions is a necessity and security measures are a key concern in the process domain [6].

Existing work applies business process anomaly detection to analyze the runtime behavior of business processes to detect and prevent attacks and security incidents, cf. [7], [8]. Unfortunately, to the best of our knowledge, the existing process anomaly detection approaches – that are capable of detecting security incidents (i.e., process runtime anomalies) – are limited in regard to their support for upcoming challenges and attack vectors. These are, for example, attacks that combine multiple process instances to harden their detection.

To verify this assumption this work investigates the state of the art in process runtime anomaly detection. By doing so it provides an overview on current process focused anomaly detection approaches, methods, and techniques – along with their advantages and limitations. Hereby it is shown that a wide range of approaches already exists but that, nevertheless, several open research gaps and questions still remain unanswered. In addition similar literature reviews are commonly available for anomaly detection work in non-process focused domains, e.g., [9]. However, these do not cover process anomaly detection publications and approaches. This probably hinders the development and research in the process anomaly detection domain because, for example, a guidance towards uncovered gaps or open questions is currently missing. Hence, we assume the given systematic literature review as necessary.

This paper is organized as follows: Section II outlines the research methodology for the conducted systematic literature review, starting with research challenges, literature search, and search term definition. The knowledge gained during this section is utilized in Section III which reports on aspects related to process anomaly detection (e.g., the nature of the analyzed data) and challenges that harden anomaly detection in the process domain. Furthermore, Section IV presents the identified anomaly detection approaches/publications – including their techniques, advantages, and disadvantages – and reveals research gaps. Limitations and gaps identified during the conducted literature analysis are outlined and discussed – in detail – in Section V. Finally, Section VI discusses the limitations of this work and Section VII concludes it.

## II. Methodology

In the following sections a systematic literature review is performed [10], [11]. To foster the soundness of the review results we follow broadly acknowledged research guidelines for systematic reviews, cf. [10]. Accordingly we start by outlining the methodology of this review, cf. Fig. 1. Initially, a preliminarily search for existing business process anomaly detection work was conducted. Because none was found research questions were formulated. Subsequently, a literature search and selection was conducted in major public research databases based on chosen broad search keywords. Identified relevant literature was analyzed, classified, and presented. Finally, research gaps and limitations were determined.

This work focuses on evaluating the anomaly detection attempts in the business process domain based on the following research questions:

**RQ1** What are anomalies and how are they defined in the process domain (common understanding and terminology)?
**RQ2** Which anomaly detection approaches are currently applied in the business process domain?
**RQ3** Which challenges and requirements remain unanswered by existing process anomaly detection research?

Each questions focuses on a different aspect which this systematic review is interested in. The first question (**RQ1**) enables to provide a general understanding of the business process anomaly detection domain (e.g., its intentions, characteristics, and goals). This requires to identify relevant work based on search keywords. To reduce the odds that some work is overlooked we opted to choose broad keywords which were combined based on boolean AND/OR operators. So during the conduced literature search the related work has to contain "*workflow*" or "*business process*" along with the term "*anomaly detection*" in its title, keywords, or abstract.

The second question (**RQ2**) focuses on determining which anomaly detection approaches are currently applied. In addition their advantages and disadvantages are identified along with our impression on the evaluation quality of the analyzed work (e.g., if the existing work is evaluated with real world data or with synthetic data). This also enabled us to determine the applied foundational anomaly detection technology (e.g., process mining or statistical techniques) which is, in the following, exploited to classify/organize the identified work.

Finally, all the findings are condensed and analyzed as a whole to identify novel research questions, limitations, and gaps (**RQ3**). Through this future research is guided.

The conducted literature search mainly focused on public research databases and libraries. Horizontal searches were conduced with Google Scholar (https://scholar.google.com/) and DBPL (http://dblp.uni-trier.de/). Furthermore, selected publisher databases were searched, i.e., IEEE Xplore (http://ieeexplore.ieee.org/), ACM Digital Library (http://dl.acm.org/), and ISI Web Of Science (https://apps.webofknowledge.com) – ISI also includes the Springer database. In total thousands of potential papers were identified in the chosen research databases. This isn't surprising as the keywords utilized during the search were broadly defined. Hence, an extensive literature selection and review process was conduced. Hereby, the title, abstract, and keywords were analyzed in a first step. This already enables to broadly classify the identified work as potentially relevant or not. Finally, the potentially relevant publications were closely examined to determine if its contents focuses on business process anomaly detection. If this is the case then the work was classified as relevant and included in this review. Finally, backward and forward snowballing was applied on the papers identified during the research database search. Overall 35 process runtime behavior anomaly detection papers, which are all listed in this paper's references, were identified as relevant and selected for the following analysis.

Be aware that in case of duplicate publications (i.e., a conference paper that was later extended into a journal article) the journal article was preferred. This is because we assume that journal articles provide more detailed and complete information when comparing them to proceeding papers. In addition we limited our search to papers that were published until (including) 2016 to foster the reproducibility of this work. This is because throughout this systematic literature review the year 2017 was not yet over so that all relevant papers of 2017 were, likely, not yet released. Hence, future reproductions of this work would likely not yield the same search results – which is taken into account by the applied approach.

### A. Publication time frame and sources

The identified literature was published between 2005 and 2016. A detailed overview on the publication time frame is provided in Fig. 2. During the early years, i.e., 2005 to 2009 a limited but nevertheless measurable increase in the publication numbers can be observed. Subsequently, between 2010 and 2014 the number of publications in each year is relatively stable, wavering between 2 to 3 for each year. Finally, a significant increase in publication numbers was observed for the years 2015 and 2016, i.e., between 5 (2015) to 10 (2016) related publications, identified by our search, were published in each of this two years. This findings indicate that the scientific community has an increasing interest in the business process runtime behavior anomaly detection domain.

In addition Table I provides an overview on the most important publication sources. This are conferences and journals that more than one of the identified business process runtime anomaly detection publications were published in. Overall it was found that the publication sources are wide spread as the identified work was found to be published in more then 28 different sources. In addition, we were left with the impression that conferences are the preferred publication method for the identified work as almost twice as much of the identified publications were published in conferences than in journals. Finally, the most significant publication source was found to be the Business Process Management (BPM) conference as a total of three publications were published in BPM proceedings.

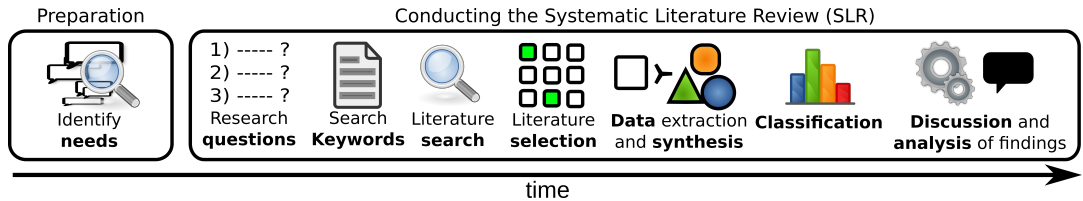

Fig. 1. Outline of the applied research methodology.

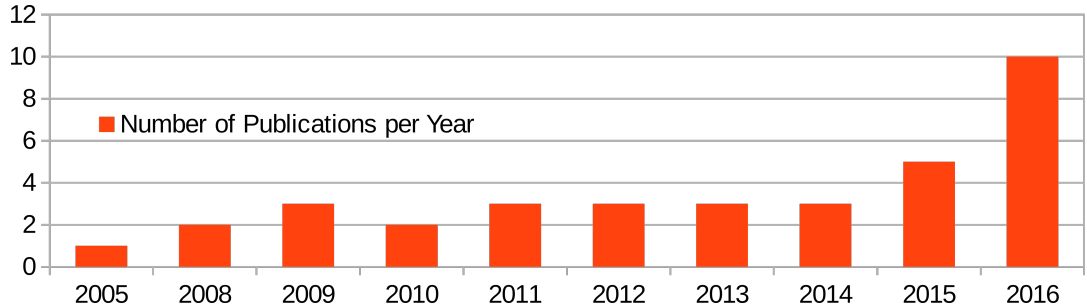

Fig. 2. Number of publications on business process runtime anomaly detection published from 2005 to 2016.

## III. Aspects and Challenges of Business Process Runtime Behavior Anomaly Detection

This section discusses different aspects and challenges of business process behavior anomaly detection – in particular – and the business process modeling domain in general. This is necessary as existing anomaly detection work concentrates and specializes on various, partly diametrical, challenges, targets, and characteristic, such as, the tackled anomaly types, the nature of the analyzed data, or domain characteristics. This shows the diversity and complexity of the domain, challenges, and requirements – which motivates not only this systematic literature review but also the application of different anomaly detection techniques and approaches.

An anomaly, in its abstracted definition, is some kind of behavior that does not conform to expected behavior, cf. [9]. Hence, anomaly detection approaches typically define the normal/expected behavior based on given (training) data. Any observed behavior that does not match to the expected behavior is assumed as anomalous. While this seems to be straightforward, several aspects make this task challenging. For example, defining normal behavior is hard, especially if the border between normal and anomalous behavior is only vaguely defined, cf. [12]. In addition attackers can adapt themselves to appear normal. Moreover, the analyzed data can contain noise or errors that are hard to distinguish from anomalies (e.g., caused by evolving process definitions) [13]. Finally, anomalies can be defined differently, depending on the application scenario and the intentions of the surrounding environment – as outlined in the following.

*1) General Anomaly Types:* Typically existing anomaly detection approaches only focus on detecting specific anomaly types. Being aware of the anomaly types that can be identified by a detection approach enables to better understand and assess its capabilities and limitations – which motivated the following outline. Generally anomalies are classified into three groups, cf. [12], i.e., point, contextual, and collective anomalies. In the following each group will be discussed, starting with point anomalies which most existing process anomaly detection work was found to be focusing on.

*Point anomalies:* Such an anomaly occurs if an individual data instance (e.g., an individual activity execution) is considered as anomalous with respect to the rest of the analyzed data (e.g., all known executions of a specific activity), cf. [9]. We assume this as the simplest form of the three described anomaly types as identifying it does not require to correlate multiple data instances (e.g., multiple activity executions). A brief example for such an anomaly can be found in typical payment transaction processes. In such a scenario an anomaly detection approach could simply monitor the *amount* data flow variable and if it exceeds a specific threshold then the current process instance is identified with a point anomaly.

*Contextual anomalies:* This kind of anomaly is sometimes also referred to as conditional anomaly, cf. [9]. It occurs if an anomalous data instance is only anomalous in a specific context. So, while business process behavior can be anomalous in a specific context the same behavior would not be identified as anomalous in a different context. Business processes hold various types of context that can be taken into consideration when searching for such anomalies, such as, the business process model itself, resources, temporal information, data along with its sources, and so on. An easy example for a contextual anomaly could be found when analyzing yearly temperature curves that are recorded from, e.g., a crop management business process. While a negative outside temperature would be anomalous in July, the same temperature would – most likely – not be anomalous in December (at least for a middle European climate, i.e., the location would be another contextual aspect that could be taken into consideration).

| Title | Type | No. |
|---|---|---|
| Business Process Management (BPM) | Conference | 3 |
| Enterprise, Business-Process and Information Systems Modeling (BPMDS) | Conference | 2 |
| Symposium On Applied Computing (SAC) | Conference | 2 |
| Information Systems Journal (ISJ) | Journal | 2 |
| Journal of Information and Data Management (JDIM) | Journal | 2 |

TABLE I
LIST OF JOURNALS AND CONFERENCES WITH MORE THAN ONE PUBLICATION.

*Collective anomalies:* Identifying collective anomalies requires to correlate multiple data instances, cf. [9]. This is because a collective anomaly only occurs if a collection of data instances is anomalous while each individual data instance is not. This can, for example, be mapped on a collection of business process instances. While each individual business process instance might seems to be unsuspicious all instances combined could still achieve a malicious goal and should be, therefore, identified as anomalous. This anomaly type is substantially harder to identify than the previously described two, i.e., it could be exploited by attackers to hide their intentions. Hence, in the following we will especially check if existing approaches support not only point anomalies but also contextual and collective anomalies.

*2) Nature of the Data and Data Sources:* The identified anomaly detection work mainly analyzes process runtime behavior. Luckily today's Process Aware Information Systems (PAIS) frequently include logging components that record and store process executions and related aspects. Hereby, not only the control flow perspective but also more complex perspectives that hold resource or temporal aspects can be fully recorded and, because of this, analyzed. In addition standardized – but also extensible – logging formats, such as, the Extensive Event Stream (XES) format are available to share, store, collect, and transmit large collections of recorded process behavior between multiple systems, cf. [14]. Nevertheless the recorded process data (i.e., each individual process instance execution trace) is not labeled out of the box.

Here *labeled* means that available anomaly detection training data (i.e., recorded process instance traces) would be correctly denoted as *normal* or *anomalous*. Such data would be of especial interest for process anomaly detection approaches to extract the difference between both data categories to identify anomalies in unseen process executions. Unfortunately a broad application of labeled data is unlikely because the labeling must typically be conducted by human experts – which requires a substantial investment in time and money, cf. [9]. In addition process executions frequently cover a substantial amount of flexibility. Hence, constructing a data set that covers all possible normal and anomalous process execution behavior characteristics is rather difficult. Dependent on the availability of labeled data (i.e., labeled normal/anomalous behavior, only labeled normal behavior, or completely unlabeled data) three different anomaly detection concepts can be applied:

*Supervised anomaly detection:* Approaches that are applying a supervised mode have the highest labeling requirements from all three approaches. They require that normal but also anomalous process executions are correctly labeled. Hereby the data can be exploited to constructs classification models that can classify any unseen process behavior into the normal or anomalous class. The applicability of such approaches is limited as accurately labeled data is hardly available and challenging to generate. Moreover, anomalous behavior is frequently relatively rare, hence, an imbalanced distribution between both data types can be observed. This further hardens the application of supervised approaches but also paves the way for semi- and unsupervised detection techniques.

*Semi-Supervised anomaly detection:* Related detection techniques only assume that data that is labeled as normal is available. This reduces the data preparation requirements and simplifies the application of related techniques. For example, as processes control critical and complex behavior, anomalies could result in severe damages so that realistic anomalous labeled data is either hard to model or so rare that it can hardly be recorded. Hence, the recorded normal behavior is exploited to construct a model (e.g., a reference process model) that represents normal behavior. Unseen behavior that does not comply to this model is identified as anomalous.

*Unsupervised anomaly detection:* Techniques that apply an unsupervised approach do not require any labeling at all and are, because of this, easily applicable in a wide range of areas. A typical assumption of such approaches is that normal process behavior is substantially more frequently occurring than anomalous process behavior. Hence, the anomalous behavior is infrequent/unlikely and can be identified by exploiting this assumption. However, if this is not the case then the anomaly detection performance drops significantly.

We found that the process anomaly detection domain mostly focuses on unsupervised and semi-supervised approaches.

*3) Challenges for Process Anomaly Detection:* Before starting with describing the different identified anomaly detection approaches, we would like to point out challenges which must be tackled by anomaly detection approaches in the

Design Time ➔ Real Time ➔ Ex Post
*high* value gained from a detected anomaly *low*

Fig. 3. Anomaly identification, points in time.

business process domain. We assume that this does not only simplify it for the reader to assess the capabilities of existing state of the art approaches but also it enables to recapitulate, in the following, which challenges are – partly – sufficiently addressed or still must be considered as future work.

**Behavior Lifecycle** Process behavior anomalies can be identified at three points in time. This is, the detection of anomalies during design time (i.e., when process models are created and their behavior is specified), real time (i.e., during the runtime/execution of a process instance/ specified behavior), or ex post (i.e., after the execution of a process instance/behavior has completed), cf. Fig. 3. We assume that the detection of an anomaly becomes less valuable for an organization the longer the timespan is between the detection and the occurrence of the anomaly, cf. [13]. Hence, detecting anomalies during design time would be optimal. However, at this point in time all the dynamic runtime behavior is not yet available (e.g., the definite value of data variables utilized during a process execution). Hence, we see real time anomaly detection as the next best alternative. In comparison the application of ex post anomaly detection can, likely, only identify that an anomaly has occurred but not prevent it as the anomalous business process instance has already completed and the anomalous behavior was fully executed.

**Change** Business processes are executed in a dynamic ever *changing environment*. Hence, to meet a range of requirements that origin from this changing environment processes are frequently adapted [15], [16]. In addition the flexibility provided by processes is exploited in day to day work life to react on individual cases which are not sufficiently supported by the existing process definitions [15]. Change and the flexibility provided by today's PAIS is diametrical to the requirements of anomaly detection approaches. This is because a stable and repetitive behavior simplifies the definition of normal behavior models and the identification of anomalous deviations.

**Interoperability** Business processes are applied in complex *inter-organizational* and *inner-organizational* scenarios [17]. In such scenarios they enable the interoperability between different systems and services with different data formats and protocols. Hence, process focused anomaly detection approaches must cope with a mixture of various runtime execution scenarios, process behavior, and data. To support this flexibility process anomaly detection work should not focus only on single selected protocols or systems – as existing anomaly detection work from other domains frequently does, cf. [9]. In addition, the available behavior data can change significantly between different PAIS. So process anomaly detection approaches are in need to support this flexibility by providing methods to take varying and fluctuating behavior data into account.

**Criticality/Communication** Business processes tend to be executed in an automatic high pace fashion, cf. [18]. Hence, when deploying an anomaly detection technique in a real world scenarios it must not only provide a sufficient computational performance but also, and likely even more important, a sufficient anomaly detection performance. This is, because automatic anomaly counter measures, such as halting or terminating process instances which were identified as anomalous, can have a significant negative impact on an organization's success. Imagine, for example, that non-anomalous business process instances are incorrectly identified as anomalous and, because of this, automatically terminated. Likely, this could hinder an organizations cooperation with its partners and the handling of customer interactions.

Hence, anomalies not only need to be identified but also assessed (e.g., harm assessment) and communicated. For this, for example, *root cause analysis* needs to be conducted to give security experts but also automatic countermeasures the necessary information to correctly decide if and how to deal with a detected anomaly. Alternative methods, such as, only alarming a security team that "something bad has happened" could result in a flooding of such teams with incorrect alarms – which could significantly reduce the trust in such a system.

**Parallelism** Business processes are executed in a highly parallel fashion. Hence, the same process definition is frequently executed simultaneously multiple times based on multiple process instances, cf. [19]. However also different business process definitions are likely executed simultaneously. This could be exploited by attackers to hide attacks. For example, an attacker could combine multiple business processes whereby each one executes only a limited portion of a larger attack. Hereby, each instance for its own could seem to be inconspicuous. Hence, when only considering each business process instance individually such an attack scenario would, most likely, not be identified. Identifying such attacks would, inter alia, require to support the detection of collective anomalies (i.e., the correlation of multiple business process instance executions is required).

**Perspectives** Business process anomaly detection approaches are in need to incorporate a wide range of different aspects into their analysis methods. This is because the behavior that must be analyzed spawns over multiple process perspectives (e.g., the control, organizational, resource, or temporal perspective). In addition process executions create, manipulate, and transformed data during their execution. Hence, process anomaly detection approaches must be defined in a flexible and extensible way to support various forms of runtime/execution behavior.

When analyzing the listed challenges and domain characteristics it becomes evident that process anomaly detection takes place in critical, flexible, and highly parallel scenarios and has to take a plethora of different perspectives, data formats, and protocols into account. The following sections are used to analyze how and if state of the art process anomaly detection approaches are coping with this situation.

## IV. BUSINESS PROCESS RUNTIME BEHAVIOR ANOMALY DETECTION APPROACHES

All identified anomaly detection approaches are presented and discussed in the following. For this the publications are structured/classified based on their foundational anomaly detection technology. Moreover, each approach and foundational technology is briefly described along with its advantages and disadvantages (including, e.g., the anomaly types the identified approach focuses on). This enables to identify gaps and open challenges and provides an overview on the process runtime anomaly detection state of the art publications.

### A. Classification based techniques

Classification based techniques typically generate/learn a classifier (e.g., a reference process model) based on given logged process execution behavior. Hereby, given process behavior is utilized as labeled – normal – training data. Hence, a common assumption of these approaches is that the classifier generation data does not contain anomalies. By comparing executions to the generated classifier they can be classified as normal or anomalous (e.g., an anomaly could be process behavior that cannot be mapped on a generated reference process based classifier). Hence, classification based techniques are applied under the assumption that a classifier can be learned from given behavior that enables to distinguish normal and anomalous process executions. In the following we discuss a number of techniques that are applied to generate classifier for business process anomaly detection purposes.

*1) Process Mining based:* Process mining was found to be a common classifier generation approach. The typical process mining focused anomaly detection approach operates in two steps. First, one of many available process mining algorithms is applied (e.g., the mining approach presented in [20]) to generate a reference process model that normal process behavior should conform to (i.e., training phase). Secondly, during the testing phase process behavior is compared with the just generated reference model, i.e., by mapping the behavior on the model. Several variants of this approach have been proposed which, for example, generate a model from a different amount of traces (e.g., the most frequent traces or 50% of all traces, cf. [2], [3], [21]) while assuming that normal behavior can be mapped to the generated model one by one, cf. [22]–[25], or with a minimal amount of model modifications (i.e., normal behavior should show structural and behavioral similarity to the classifier model, cf. [7]). Similar approaches are proposed in [26]–[28] where the authors identify frequent traces, create a model from these traces, and calculate the conformance of infrequent traces to this model to identify outliers/anomalies. Especially [29] is interesting in this area. This is because this work couples [28] with a map reduce based approach to enable a parallelized analysis of large process logs.

Another rather unique approach is proposed in [30]. This work mines an automate that represents all behavior and subsequently removes/cleans infrequent behavior from it. Behavior that can no longer be mapped on the automate after cleanup is identified as anomalous. This approach can be computational intense because it identifies an "optimal" way to clean up the automate in multiple iterations. The work in [31] proposes a window based approach that slides over the process behavior and calculates an anomaly score for each window based on its exact reproducibility frequency in all the logged behavior.

We found that the main advantage of process mining based approaches is their deep rooting in the process domain. Hence, they support a wide range of process characteristics and are specifically crafted to deal with related challenges and data (e.g., logged execution behavior with varying data quality). In addition if was found that existing process mining approaches especially focus on the control flow perspective. This also affects the listed anomaly detection work, i.e., non control flow related process perspectives are currently not observed for anomalies by these approaches – which limits their anomaly detection capabilities. In addition the listed approaches only support point anomaly detection. We assume this as a frequently occurring substantial limitation which was not only observed for these work but also for most of the anomaly detection work discussed in the following.

*2) Rule Mining based:* Rule based anomaly detection techniques strive to learn and identify rules that represent normal process execution behavior. Unseen behavior can subsequently be compared to these rules and if the rules are not matching then the behavior is considered as anomalous. In addition reversed approaches are also conceivable. Hence, known anomalous behavior is utilized to construct rules that match only to anomalous but not to non-anomalous process behavior.

Comparable to process mining, rule mining also applies a two pronged approach. Hence, first, existing rule mining techniques, such as, association rule mining or decision trees are applied. Subsequently, classic anomaly detection work typically assigns an anomaly score to each rule (e.g., to determine how important a rule is, i.e., not all rules must match). Most process anomaly detection approaches were found to apply a stricter approach so that even minimal violations to the applied rules classify business process behavior as anomalous.

Several techniques have been proposed for process anomaly detection rule mining. For example, Support Vector Machine techniques are applied in [32] to mine rules which must be strictly met by a processes to not be considered as anomalous. A different approach is applied by the work in [33]–[35]. These work focuses more on detecting rules which, when met, identify anomalous behavior. For this, explorative process mining is applied by experts to generate a labeled training set. The approach in [35] includes ontologies to gain additional knowledge about the processes under analysis.

The main advantage of the proposed rule mining approaches is that they generate a, probably small, set of rules that likely enables to differentiate anomalous from non-anomalous process behavior. However, the strictness enforced by the identified approaches likely increases the change for incorrect analysis results in the case of flexible or evolving process definitions. Moreover, a substantial part of the identified rule based approaches is heavily dependent on manual interventions and expert knowledge. Hence, experts must mark whole execution traces as normal or anomalous (i.e., to generate labeled training data) and review generated rules/combination of rules to achieve a good anomaly detection performance.

Some of identified rule mining approaches, e.g., the work in [33], were found to support contextual and point anomalies along with multiple perspectives. However, the given support for perspectives other than the control flow left the impression of being overly simplistic (e.g., temporal aspects were supported but only by a manually generated global rule that each activity duration must not exceed 15 minutes).

*3) Clustering based:* Clustering based techniques, such as, nearest neighbor, were found to be the most commonly used process anomaly detection work. These work assumes that normal process behavior "occurs" in dense neighborhoods. Hence, while normal process executions will be packed close to each other anomalous ones will be far from their closest neighbors or build an independent suspiciously small cluster.

This requires to measure the distance or similarity between multiple process executions. For this [36] converts the behavior of resources into vectors to measure the distance between these vectors. Hereby, the grounding assumption is that resources that are assigned to the same roles should behave similar. Similarly [37] focuses on temporal aspects, i.e., it is assumed that the same activity behaves similar in all executions with regard to their temporal aspects (e.g., its execution time). Finally, [38] focuses specifically on the control flow. For this patterns are mined and utilized to cluster process instances (e.g., based on the supported control flow patterns).

More control flow focused techniques are presented in the following. Hereby, basically, the distance between an expected and an observed activity sequence is calculated (e.g., based on longest common subsequence mining). If this distance becomes, for example, to high (i.e., higher than a manually defined threshold) then an anomaly was found [39]–[43]. A different technique is proposed in [44], which utilizes a limited set of patterns whose likelihood is calculated. This enables to briefly communicate why an anomaly was detected, e.g., because specific patterns were missing or occurred to early/ were delayed. Unfortunately, the patterns must be manually created which requires a substantial amount of domain and expert knowledge along with up-to-date documentation.

Finally, [13] proposes an unsupervised clustering/likelihood based approach. Hence, a reference model is generated that enables to calculate the likelihood of process executions. If an unseen process instance is unlikely, compared to other executions of the same process, then the instance is assumed as anomalous. Especially [13] sicks out compared to the other work because it implements an extensible approach that has shown how the control, resource, and temporal perspective can be supported at once.

Some of the presented clustering based techniques were found to incorporate contextual attributes. For example, [37] analyzes activity durations with respect to an artificial assumed resource experience, i.e., [37] assumes that, e.g., advanced clerks process requests faster. Herby the detection quality can be improved and non control flow related aspects are taken into consideration. Still these techniques frequently only focus on point anomalies, e.g., single activity durations are analyzed (except [13] which is also capable of detecting collective anomalies). This could be exploited by inside attackers that do not rush through the processes but instead show a normal execution performance. In addition the similarity and distance calculation can be computational intense, for example, the pattern detection in [38] can be generalized to sub sequence mining which is assumed as NP hard.

*4) Neural Network based:* Neural network based techniques support a wide range of application areas. For this a neural network is typically trained to differentiate between normal and anomalous process executions. However, the identified work in [45] applies a different approach. The authors decided to train the network so that it can reproduce the input trace that should be analyzed. If the input and the reproduced trace are not similar enough (again using a manually defined threshold) then an anomaly was found. The applied network has a fixed input layer size. Hence, traces could be too large to be handled by the network which, in the case of, e.g., loops, could be exploited to render this approach inapplicable which, e.g., enables to hide attacks. In addition this approach only focuses on point anomalies and the control flow perspective.

*5) Support Vector Machines based:* Another classification approach are Support Vector Machines (SVM). Typically anomaly detection approaches apply SVM to partition the data space into regions that contain the analyzed data instances. If an unseen data instance (i.e., process behavior) does not fall in one of the predefined regions then an anomaly was found. In comparison [8] uses the regression calculation capabilities of SVM to calculate the most likely destination airport for a plane based on position, elevation, speed, and so on – these information is collected by monitoring processes. Overall [8] focuses less on process anomalies than the other work presented here. Nevertheless it is an interesting extension of this field because it analyzes data variables collected by business processes executions – which is rarely the case in the identified process anomaly detection work.

*6) Regular Expression based:* Regular expression based techniques define classifiers as regular expressions. Hence, training data is analyzed to learn a regular expression that matches this data. Data that does not match to this regular expression is, accordingly, assumed as anomalous. The approach presented in [46] applies this concept to reason about data that is exchanged during process model executions (e.g., between activities or process partners). In addition this work proposes to support contextual anomalies based on generating different regular expressions for different contexts (e.g., resources or processes that the data to analyze originates from). While this provides a flexible way to analyze data it can be computational intense to calculate and evaluate the regular expressions. In addition if normal data is fluctuating (e.g., the data can be defined in different orders without changing its meaning) then this flexibility can hardly be covered by "static" regular expressions without resulting in overly complex expressions.

*B. Statistics based techniques*

Statistical anomaly detection techniques typically generate a stochastic model that represents analyzed behavior. If unseen behavior is mapped on the low probability region of this stochastic model then an anomaly was found. For this a statistical model is fitted (i.e., trained) to the given normal data. Such an approach is applied in [47] to detect temporal anomalies (i.e., unexpected durations of activities as point anomalies). Statistical approaches are heavily dependent on a fine tuned and correct assumption of the data distribution used to generate the stochastic model. Getting this right might be simple for a limited set of dimensions but gets substantially harder for multi dimensional data (i.e., if multiple process perspectives should be incorporated).

*C. Other approaches*

The previous sections concentrate on process behavior analysis approaches that are capable of detecting anomalous behavior in an automatic fashion. However, different approaches are also conceivable. For example, [8] proposes multiple behavior visualization techniques to ease the visual inspection and anomaly detection of process execution behavior for experts. Coupling this approach with previously discussed techniques could potentially simplify the root course analysis of an automatically detected anomaly. Other approaches, such as the ones presented in [48], [49] focus more on the behavior definition side (i.e., process design time). Hence, they enable – based on predefined patterns – to detect anomalies, for example, in the data flow perspective of a process model (e.g., when data is read without being written before).

Overall it can be concluded that the process anomaly detection domain heavily focuses on classification based techniques, in general, and process mining, in particular. When comparing this to the overall state of the anomaly detection domain (e.g., including network security or hardware maintenance focused anomaly detection, cf. [9]) then this picture seems to be unbalanced as, e.g., statistical approaches are underrepresented. Hence, it would be a promising research direction to explore the applicability of those approaches in the process anomaly detection domain – which motivated the following section where this and other gaps are discussed.

## V. Open research gaps and directions

This section exploits the findings of this systematic literature review to identify open research gaps in the process runtime anomaly detection domain. For this, Fig. 4 depicts a summarized overview on our findings. For example, it shows the number of publications that concentrate on each anomaly type, process perspective, or that apply a specific fundamental technology. Moreover, it depicts if and with which data the identified approaches were evaluated and if the identified work is focusing on design time, real time, or ex post anomaly detection. If possible a publication was counted multiple times (e.g., if a publication covers point but also contextual anomalies or applies real world and synthetic evaluation data).

*a) Detecting anomalies early:* Today's process runtime anomaly detection approaches were found to be frequently focusing on *ex post* analysis, cf. Fig. 4. So, they require that, for example, a process instance was fully executed and completed before the related behavior can be analyzed for anomalies. However, at this point in time fraudulent behavior or a malicious intention was already fully executed/achieved. Hence, more flexible anomaly detection approaches are required that can also analyze incomplete process executions to detect anomalies early, i.e., before the executed of an anomalous business process instance has completed.

*b) Root cause analysis/Communication:* We found that the identified process behavior runtime anomaly detection approaches typically operate in a *black box mode*. So, for example, they either only present to the user which business process executions were detected as anomalous or, the more communicative techniques, also provide an anomaly score. This score indicates how much detected anomalous behavior deviated from the expected behavior. This is likely insufficient to truly explain the nature and cause of an anomaly to an end-user. However a detailed explanation, likely, is important to *a)* truly assess the severeness of the anomaly by security experts; and *b)* pick an appropriate counter measure (e.g., halting or terminating the process instance in question). Currently the most promising approaches in this area are the identified rule mining techniques. This is because those can output the rules that motivated the normal or anomalous decision.

*c) Looking beyond point anomalies:* It was found that most existing process anomaly detection work focuses on rather simple anomaly types, i.e., *point anomalies*. In addition a very small amount of all found publications propose approaches that take contextual aspects into account – these approaches were found to mainly focus on the temporal and resource *perspective*, cf. Fig. 4. Even publications that cover more than one perspective mostly only bundle two of the multiple available process perspectives in a rather "closed" way that can, based on our impression, hardly be extended to take additional perspectives into account (a majority of the found work even only supports the control flow perspective). This results in gaps and unmonitored areas which could be exploited by attackers to hide their intentions. To address this limitation approaches must be constructed that can incorporate multiple perspectives at once and that can be dynamically extended to take new perspectives and developments into account. As a quick solution it could be tried to aggregate the anomaly scores generated by multiple point anomaly detection approaches. Nevertheless this still leaves collective anomalies behind which are, based on our findings, currently hardly covered by almost all process anomaly detection work.

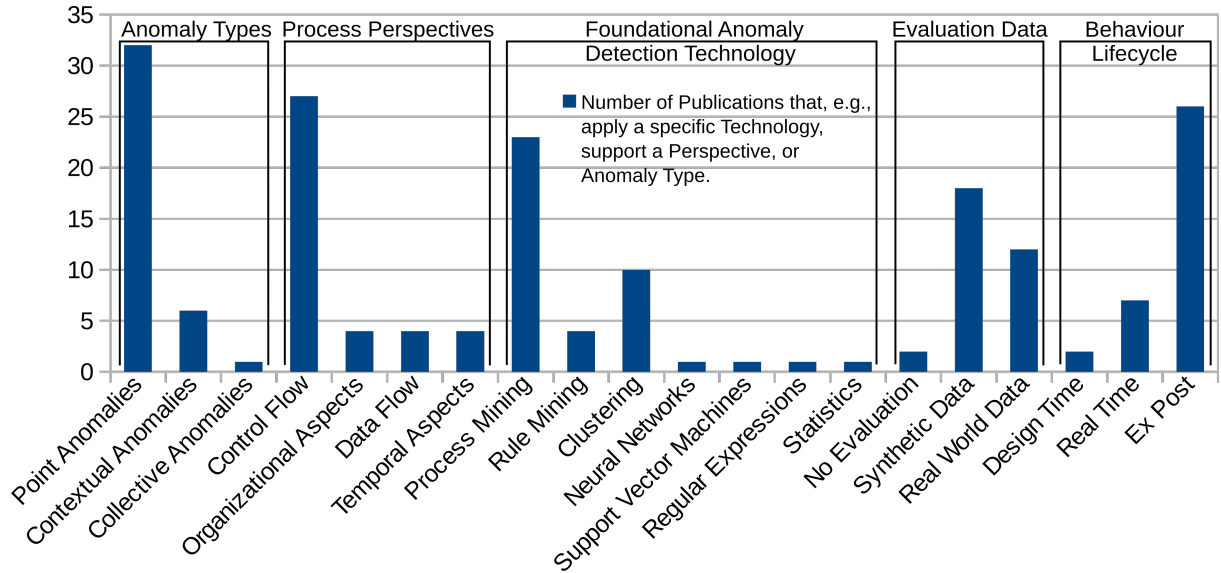

Fig. 4. Quantitative overview, e.g., on applied techniques, supported anomaly types, and process perspectives.

*d) Interoperability and Parallelism:* Processes behavior takes place in complex *parallel*, *inter-*, and *inner-organizational* scenarios and environments. However, it was found that only non-parallelized inner-organizational behavior is taken into consideration by the existing process anomaly detection work. Hence, neither did the identified publications support the monitoring of multiple parallel process instances (from the same or different models) nor inter-organizational process execution behavior. Through this typical application areas are not yet secured and attackers could, for example, split up attacks in multiple process instances to hide them. We assume this situation as a critical gap which must be addressed to tackle related attack vectors that could be exploited, e.g., to hide attacks from existing anomaly detection approaches.

*e) Handling change and flexibility:* Most of the analyzed work were found to be likely struggling with flexible, *changing*, and *evolving* business process definitions. Hence, the listed approaches typically expect fixed and stable process behavior whereby minimal deviations from, for example, an expected path are already sufficient to conclude that an anomaly has occurred. We assume that such approaches are too strict because processes are evolving around the clock to meet new requirements, cf. [15], [16]. In addition the flexibility provided by today's PAIS is well received and used to deal with non-standard cases. However, these flexibility/constant change is currently hardly supported which can result in assessing correct behavior as anomalous. In addition it can result in a frequent need to recalculate, for example, anomaly detection classifiers – which can be computational intense – because they cannot cope with the evolution of the underlying business process models. This situation does not only affect the runtime behavior that is analyzed for anomalies but also the exploited data source, i.e., process execution logs. Hence, robust approaches are necessary, but hardly available, to deal with noise or partly anomalous process execution logs. For example, incremental rule mining could be a promising starting point to address this situation, but was not yet explored.

*f) Unbalanced techniques:* Finally, as already pointed out previously, the techniques applied in the business process anomaly detection domain were found to be *unbalanced*, cf. Fig. 4. We assume that this unnecessarily neglects countless research efforts and significant achievements in other domains. These likely provide valuable results and starting points that can be extended to be applicable to the process domain.

*g) Evaluation quality:* Overall we found that a significant part of the analyzed work shows only a *limited evaluation quality*, cf. Fig. 4. This is, that most work were found to either contain no evaluation or that the evaluation was only conducted with synthetic data (e.g., synthetic process execution logs). Hence, we assume it as an important aspect to ensure, e.g., the reproducibility of the proposed approaches under real world conditions. This requires to encourage organizations to publish real world business process behavior data, in an optimal case even known anomalous process instances that were identified by an organization's security team.

## VI. LIMITATIONS OF THIS REVIEW

This section discuss limitations and potential threats to validity of this systematic literature review.

*a) Excluded areas:* This systematic review concentrates on discovering and analyzing process runtime behavior anomaly detection publications. Through this non process runtime anomaly detection focused approaches, such as the briefly outlined approaches in Section IV-C, were not aimed on, e.g., this area can be covered based on future work.

*b) Limitation of the search:* Due to the wide-spread utilization of anomaly detection it became necessary to limit the search (cf. Section II) to specific sections of the searched publications (title, abstract, and keywords). In addition we concentrated with this work on electronic literature databases and do not conduct, by purpose, a manual search of selected conference proceeding or journals. This is because during our previous systematic literature reviews (e.g., our work in [50]) we found that a manual search hardly identified work which was not also retrieved by the database based search. Moreover, this work includes more than six different literature databases and applies snowballing. Therefore, we are confident that we cover the most important sources and publications.

*c) Presented ideas and algorithms:* This systematic review elaborates the ideas, algorithms, and foundations of the discovered publications. However, an extensive elaboration would go beyond the scope of this study. Hence, we would like to refer the interested reader to the cited literature.

## VII. Conclusion

This systematic literature review provides a holistic view on business process anomaly detection approaches that can be applied to detect and prevent related security incidents. For this 35 publications were examined and analyzed which also enabled to identify gaps and limitations in this domain. Related to our research questions we found that:

**RQ1** The process anomaly detection domain applies an anomaly definition which is also used in other domains. For example, that an anomaly is some king of unlikely behavior or infrequently occurring behavior. Hereby, the most significant difference between the process behavior anomaly detection domain and other anomaly detection domains is that the first one operates on a specific type of data (i.e., execution traces) and faces unique challenges which we assume as rather unusual for other domains.

**RQ2** Regarding the applied anomaly detection approaches we found a mixed picture. This is, because the found business process anomaly detection work applies an unbalanced set of techniques and focuses heavily on a few specific methods and approaches. We assume that this neglects the potential of alternative anomaly detection approaches which are already applied in other domains.

**RQ3** This work was capable of identifying multiple challenges which are not yet sufficiently tackled. Hereby, we anticipate limitations when dealing with flexible and evolving process model definitions as the most important one. This is, because it was found that processes are under the constant pressure to change – this is also one of process's big advantages because they are typically capable to cope with this pressure, cf. [16]. Not supporting this key advantage with existing anomaly detection approaches, likely, limits their applicability. Moreover, most of the identified work focused solely on point anomalies leaving more complex collective and contextual anomalies undetected. This likely results in unprotected attack vectors which are not covered by existing work.

In future work, we plan to address the identified limitations to strengthen the business process domain by preventing security incidents through comprehensive anomaly detection approaches. Hence, anomaly detection approaches will be proposed that enable to detect collective anomalies in the process data flow, i.e., especially in the data exchanged between process activities and process partners – which seems to be a mostly neglected area, at least based on our findings. However, because malicious data can be the source of attacks we assume it as necessary to also detect complex anomalies in this area. Further, we plan to propose approaches to detect attacks were the attackers did split their malicious intentions into multiple process executions. Each individual one might seems to be normal but all executions combined would then achieve a malicious goal. This is hardly detectable by existing approaches because of their strong – almost exclusively – focus on point anomalies and individual process executions.